\documentclass[aps,pra,twocolumn,showpacs,floatfix]{revtex4-2}
\usepackage{graphicx}
\usepackage{subfigure}
\usepackage[intlimits]{amsmath}
\usepackage{amssymb}
\usepackage{amsfonts}
\usepackage{natbib}
\usepackage[utf8,decmulti]{inputenc}
\usepackage{hyperref}
\usepackage{cleveref}
\usepackage{array}
\usepackage{dcolumn}
\usepackage{longtable}
\usepackage{booktabs}
\begin{document}
\title{Propagation of the angular spectrum of electromagnetic fields in uniaxial crystals of finite length in the regime of the paraxial approximation}
\author{A. G. da Costa Moura}
\email{alex.gutenberg@ict.ufvjm.edu.br}
\affiliation{Instituto de Ci\^{e}ncia e Tecnologia, Universidade Federal
dos Vales do Jequitinhonha e Mucuri, Rodovia MGT 367 - Km 583, 5000, Alto da Jacuba, Diamantina, MG 39100-000,
Brazil}
\date{\today}
\begin{abstract}
In this paper we will analyze the propagation of the angular spectrum of the electromagnetic field through a finite length uniaxial crystal. We solve the boundary conditions of the fields at an interface of an isotropic medium and an uniaxial anisotropic medium with the optical axis in an arbitrary direction and soon after we choose the optical axis in a plane formed by the direction of propagation (z-axis) and the x-axis . We show how the couplings occur between the components of the field and in what situations we can give a vector or scalar treatment for the propagation.
\end{abstract}
\pacs{41.20.Jb, 42.25.Lc, 42.30.Kq, 42.25.Db}
\maketitle
\section{Introduction}
Over the years, many works on the propagation of electromagnetic fields in birefringent media have been published\cite{yang,qi,ziren,xie}.
The mathematical tool widely used for the treatment of this propagation is Fourier Optics, which consists in writing the electromagnetic fields in a superposition of plane waves, which allows to simplify the analysis of the propagation of the fields.
\begin{figure}[h]
  \includegraphics[width=4.0cm]{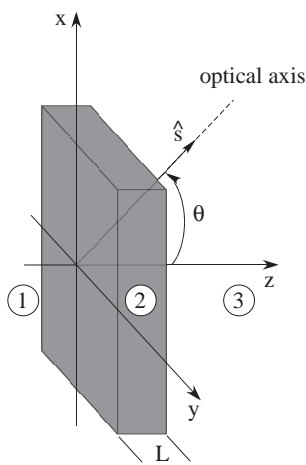}
  \caption{\label{cris}Uniaxial crystal. The incident field (media $1$) enters the crystal, propagates through this media (media $2$) whose optical axis is oriented in the $xy$ plane, forming an angle $\theta$ with the direction $z$, and emerges on the right side of the crystal (media $3$). Media $1$ and $3$ are isotropic. The direction of the optical axis is represented by the unit vector $\hat{s}$.}
\end{figure}
Some authors have given a vector treatment to propagation within uniaxial crystals, analyzed the transmission and reflection in a plane interface between an isotropic and a uniaxial medium \cite{ciattonipalma, ciattonidiporto, ciattonicjpal} but all these works considered the propagation of the fields in the direction of the optical axis or perpendicular to it.
The representation of monochromatic electromagnetic beams in an isotropic half-space by plane waves has been used with great success for several decades. When we treat propagation in isotropic media, Fourier Optics is a scalar theory valid for any of the components of the field. The situation changes completely when propagation occurs in anisotropic media in which the propagation of the electromagnetic fields depends strongly on polarization. In this case, the concept of angular spectrum remains valid, although it is no longer a scalar quantity \cite{lalor}.
In this work we present results that derive from a general theory developed by Stamnes et al., who made a very general formulation for the propagation of electromagnetic beams in anisotropic media
\cite{stamneslalor,stamnessherman76,stamnesgasper76,stamnessherman77,stamnesdaya98}.
In this paper we describe the angular spectrum propagating through a uniaxial crystal of length L, as shown in figure \ref{cris}, and the effects of the anisotropy in the field transmitted to the medium 3. We determine a transfer matrix that informs the diffraction and the coupling of fields . This transfer matrix appears due to the boundary conditions at the crystal interface.
\section{The polarizations of the field and the paraxial approximation}
In an anisotropic medium, the phase velocity of a plane wave depends on the direction of its wave vector $\vec{k}$ and its polarization state. In the uniaxial media there is a direction, called the optical axis, for which the phase velocity is independent of the polarization state of the wave. For all other directions of the wave vector, there are two polarization auto-states known as ordinary polarization and extraordinary polarization, which correspond to two different phase velocities. For each polarization auto-state we have a index of refraction: $n_ {o}$ for ordinary polarization and $n_e$ for extraordinary polarization. Here we will assume that the optical axis will be in the plane $xz$ and makes an acute angle $\theta$ with the positive z-axis ($-90^{\circ}<\theta<90^{\circ}$). In order for the plane waves in the uniaxial medium to be compatible with the Maxwell's equations, the components of the wave vectors must satisfy \cite{wolf}
\begin{equation}
\frac{k_{x}^{2}+k_{y}^{2}+k_{z}^{2}}{n_{o}^{2}}=\frac{\omega^{2}}{c^{2}}
\label{eq:esfera}
\end{equation}
and
\begin{eqnarray}
&&k_{x}^{2}\left(\frac{\cos^{2}\theta}{n_{e}^{2}}+\frac{\sin^{2}\theta}{n_{o}^{2}}\right)+\frac{k_{y}^{2}}{n_{e}^{2}}+
k_{z}^{2}\left(\frac{\cos^{2}\theta}{n_{o}^{2}}+\frac{\sin^{2}\theta}{n_{e}^{2}}\right)\nonumber\\
&&+2k_{x}k_{z}\left(\frac{1}{n_{e}^{2}}-\frac{1}{n_{o}^{2}}\right)=\frac{\omega^{2}}{c^{2}}
\label{eq:elipse}
\end{eqnarray}
for ordinary and extraordinary polarizations, respectively.
To specify the directions of polarization in the uniaxial medium, we will adopt the following definitions:
$\hat{s}$ is a unit vector in the direction of the optical axis; $\vec{k}^{o}$ is the wave vector of a plane wave with ordinary polarization; $\vec{k}^{e}$ is the wave vector of a plane wave with extraordinary polarization; $\hat{k}^{o}$ and $\hat{k}^{e}$ are unit vectors in the directions of $\vec{k}^{o}$ and $\vec{k}^{e}$, respectively; $\hat{\epsilon_{o}}$ is a unit vector in the direction of ordinary polarization;
$\hat{\epsilon_{e}}$ is a unit vector in the direction of extraordinary polarization.

It is possible to show that \cite{lalor,stamnessherman76,stamnessherman77}
\begin{eqnarray}
\hat{s}&=&\sin\theta\hat{x}+\cos\theta\hat{z},\\
\vec{k}^{o}&=&n_{o}\frac{\omega}{c}\hat{k}^{o},\\
\vec{k}^{e}&=&\frac{n_{o}n_{e}}{\sqrt{n_{o}^{2}\left|\hat{k}^{e}\times\hat{s}\right|^{2}+n_{e}^{2}
\left(\hat{k}^{e}\cdot\hat{s}\right)^{2}}}\frac{\omega}{c}\hat{k}^{e},\\
\hat{\epsilon}_{o}&=&C^{o}\vec{k}^{o}\times\hat{s},\\
\hat{\epsilon}_{e}&=&C^{e}\left[\left(\vec{k}^{o}\cdot\vec{k}^{o}\right)\hat{s}-
\left(\vec{k}^{e}\cdot\hat{s}\right)\vec{k}^{e}\right],
\end{eqnarray}
where
\begin{eqnarray}
C^{o}&=&\frac{1}{\left|\vec{k}^{o}\times\hat{s}\right|},\label{co}\\
C^{e}&=&\frac{1}{\left|\vec{k}^{e}\times\hat{s}\right|\sqrt{n_{o}^{4}\left|\vec{k}^{e}\times\hat{s}\right|^{2}+n_{e}^{4}
\left(\vec{k}^{e}\cdot\hat{s}\right)^{2}}}\label{ce}.
\end{eqnarray}
The angular spectrum of the beam will have two components that correspond to the expansions in plane waves with ordinary polarizations, $\mathcal{E}^{o}$, and extraordinary polarization, $\mathcal{E}^{e}$, in the form
$\vec{\mathcal{E}}=\vec{\mathcal{E}}^{o}+\vec{\mathcal{E}}^{e}$. The transverse component of $\vec{k}^{o}$ and $\vec{k}^{o}$ is $\vec{q}=k_{x}\hat{x}+k_{y}\hat{y}$.
We can write the electric field E in the form
\begin{equation}
\vec{E}(\vec{r})=\int_{\mathbb{R}^{2}}\mathcal{E}^{e}\hat{\epsilon}_{e}e^{i\vec{k}^{e}\cdot\vec{r}}d\vec{q}+
\int_{\mathbb{R}^{2}}\mathcal{E}^{o}\hat{\epsilon}_{o}e^{i\vec{k}^{o}\cdot\vec{r}}d\vec{q}.
\end{equation}
All terms within integrals are functions of $q$, except $r$.
From equations (\ref{eq:esfera}) and (\ref{eq:elipse}), we can show that
\begin{eqnarray}
\vec{k}^{o}&=&\vec{q}\pm \sqrt{k^{2}-q^{2}}\hat{z},\label{eq:kord}\\
\vec{k}^{e}&=&\vec{q}\pm \left(\alpha q_{x}+\sqrt{\kappa^{2}-\beta q_{x}^{2}-\gamma q_{y}^{2}}\right)\hat{z},\label{eq:kext}
\end{eqnarray}
where the $+$ ($-$) sign indicates propagation towards the positive (negative) z axis, and
\begin{eqnarray}
k&=&n_{o}\frac{\omega}{c},\label{kno}\\
\kappa&=&\eta\frac{\omega}{c}\label{kappa},\\
\eta&=&\frac{n_{o}n_{e}}{\sqrt{n_{o}^{2}\sin^{2}\theta+n_{e}^{2}cos^{2}\theta}},\\
\alpha&=&\frac{\left(n_{e}^{2}-n_{o}^{2}\right)\sin\theta\cos\theta}{n_{o}^{2}\sin^{2}\theta+n_{e}^{2}cos^{2}\theta},\\
\beta&=&\left(\frac{n_{o}n_{e}}{n_{o}^{2}\sin^{2}\theta+n_{e}^{2}cos^{2}\theta}\right)^{2},\\
\gamma&=&\frac{n_{o}^{2}}{n_{o}^{2}\sin^{2}\theta+n_{e}^{2}cos^{2}\theta}.\label{gamma}
\end{eqnarray}
Here, we will adopt the paraxial approximation and write the equations (\ref{eq:kord}) and (\ref{eq:kext}) in the form
\begin{eqnarray}
\vec{k}^{o}&\approx&\vec{q}\pm\left(k-\frac{q^{2}}{2k}\right)\hat{z},\\
\vec{k}^{e}&\approx&\vec{q}\pm\left(\alpha q_{x}+\kappa-\frac{\beta q_{x}^{2}+\gamma q_{y}^{2}}{2\kappa}\right)\hat{z}.\label{eq:aproparke}
\end{eqnarray}
In equation (\ref{eq:aproparke}), the $\beta$ and $\gamma$ quantities are, respectively, magnifications in the $x$ and $y$ components of the transverse moment.
The $\kappa$ quantity is the modulus of the wave vector with anisotropy, $\eta$ is the index of refraction in the direction of propagation of the extraordinary field and $\alpha$ is the term of walk-off. This term is fundamental in the transfer of the angular spectrum of an incident beam to the state of two photons generated in a parametric process and detected in coincidence \cite{alexmonken}.
\section{The transfer matrix}
To determine the field leaving a uniaxial crystal, after traveling a distance $L$,we will have to calculate the field that is transferred into the crystal. For this, we must express the spectral amplitudes outside the crystal and then, using the boundary conditions for the fields $\vec{E}$ and $\vec{H}$, calculate the field inside the uniaxial medium.
Propagating the fields inside the crystal, we must find the expressions of the ordinary and extraordinary fields and soon after, again using the boundary conditions at the interface, we will determine the field at the exit of the crystal. The angular spectra of the incident ($i$), reflected ($r$) and transmitted ($t$) electric fields will be written in the form
\begin{equation}
\vec{\mathcal{E}}^{m}=\mathcal{E}^{em}\hat{\epsilon}_{e}^{m}+\mathcal{E}^{om}\hat{\epsilon}_{o}^{m}
\end{equation}
where $m=i$, $r$, or $t$.
For the magnetic fields, we will have an analogous expression:
\begin{equation}
\vec{\mathcal{H}}^{m}=\vec{\mathcal{H}}^{em}+\vec{\mathcal{H}}^{om}
\end{equation}
Let us determine the transfer matrix that relates the fields inside and outside the uniaxial medium, initially considering two uniaxial media, media $1$ e $2$, with the optical axes oriented in the directions $\hat{s}_{1}$ and $\hat{s}_{2}$. We will treat the isotropic media as uniaxial media at the limit $n_{e}\rightarrow n_{o}\rightarrow 1$.
We will call the interface between the medias $1$ and $2$ of interface $I$ and the interface between medias $2$ and $3$ of interface $II$. The tangential components of $E$ and $H$ in interface $I$ are continuous\cite{jackson}. Considering that each component of the plane wave must satisfy the boundary conditions in the interface individually, we will have
\begin{equation}
\hat{z}\times\left[\vec{\mathcal{E}}^{i}+\vec{\mathcal{E}}^{r}-\vec{\mathcal{E}}^{t}\right]_{z=z_{0}}=0
\label{zxE}
\end{equation}
and
\begin{equation}
\hat{z}\times\left[\vec{\mathcal{H}}^{i}+\vec{\mathcal{H}}^{r}-\vec{\mathcal{H}}^{t}\right]_{z=z_{0}}=0.
\label{zxH}
\end{equation}
To facilitate calculations, it is more convenient to express the vector $\hat{s}$, which specifies the optical axis, in terms of a base associated with the transverse component $\vec{q}$ of the wave vector, as follows
\begin{equation}
\hat{s}^{m}=\sigma^{m}\hat{z}+a^{m}\hat{a}+b^{m}\hat{b}
\label{eixopt}
\end{equation}
where
\begin{eqnarray}
\hat{a}&=&\frac{\vec{q}}{q}\\
\hat{b}&=&\hat{z}\times\hat{a}.
\end{eqnarray}
\begin{figure}
  \begin{center}
  \includegraphics[width=6.0cm]{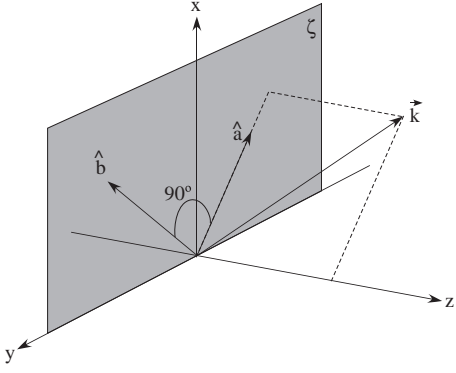}\\
  \caption{System representing the base formed by the unit vectors $a$, $b$ and $z$. The vectors $a$ and $b$ are in the plane $\zeta$, even plane $xy$.}\label{baseab}
  \end{center}
\end{figure}
If $m=i$ or $r$, equation (\ref{eixopt}) refers to medium $1$, whereas if $m=t$, it refers to medium $2$. Thus, $\hat{s}^{i}=\hat{s}^{r}=\hat{s}_{1}$, $\hat{s}^{t}=\hat{s}_{2}$, the same is true for $a^{m}$, $b^{m}$ and $\sigma^{m}$. Figure \ref{baseab} illustrates the plane that holds the unit vectors $\hat{a}$ and $\hat{b}$.
Making the scalar product of equations (\ref{zxE}) and (\ref{zxH}) for $\hat{a}$ and $\hat{b}$, respectively, we will arrive at a set of four equations for the interface I:
\begin{equation}
\left[
  \begin{array}{cccc}
    A^{or} & -D^{er}\left(k^{or}\right)^{2} & -A^{ot} & D^{et}\left(k^{ot}\right)^{2}\\
    -D^{or}k_{z}^{or} & B^{er} & D^{ot}k_{z}^{ot} & -B^{et}\\
    D^{or}\left(k^{or}\right)^{2} & A^{er}\left(k^{or}\right)^{2} & -D^{ot}\left(k^{ot}\right)^{2} & -A^{et}\left(k^{ot}\right)^{2}\\
    -B^{or} & -D^{er}\left(k^{or}\right)^{2}k_{z}^{er} & B^{ot} & D^{et}\left(k^{ot}\right)^{2}k^{et}_{z}\\
  \end{array}
\right]
\nonumber
\end{equation}
\begin{equation}
\times
\left[
  \begin{array}{c}
    \mathcal{E}_{I}^{or}\\
    \mathcal{E}_{I}^{er}\\
    \mathcal{E}_{I}^{ot}\\
    \mathcal{E}_{I}^{et}\\
  \end{array}
\right]=
\mathcal{E}_{I}^{oi}\left[
                      \begin{array}{c}
                        -A^{oi} \\
                        D^{oi}k_{z}^{oi} \\
                        -D^{oi}\left(k^{oi}\right)^{2} \\
                        B^{oi} \\
                      \end{array}
                    \right]
\nonumber
\end{equation}
\begin{equation}
+
\mathcal{E}_{I}^{ei}\left[
                      \begin{array}{c}
                        D^{ei}\left(k^{oi}\right)^{2}\\
                        -B^{ei}\\
                        -A^{ei}\left(k^{oi}\right)^{2}\\
                        D^{ei}\left(k^{oi}\right)^{2}k_{z}^{ei}\\
                      \end{array}
                    \right]
\label{matrixtrans}
\end{equation}
where
\begin{eqnarray}
A^{pm}&=&C^{pm}\left(\sigma^{m}q-a^{m}k_{z}^{pm}\right),\\
B^{pm}&=&C^{pm}\left[a^{pm}\left(k^{om}\right)^{2}-q\left(\vec{k}^{pm}\cdot\hat{s}^{m}\right)\right],\\
D^{pm}&=&C^{pm}b^{m},
\end{eqnarray}
with $p=e,o$, e $m=i,r,t$. The quantity $C^{pm}$ will be given by (\ref{co}) and (\ref{ce}). Rewriting matrix equation (\ref{matrixtrans}) in the form
\begin{equation}
\left[
  \begin{array}{cccc}
    \Omega_{11} & \Omega_{12} & \Omega_{13} & \Omega_{14} \\
    \Omega_{21} & \Omega_{22} & \Omega_{23} & \Omega_{24} \\
    \Omega_{31} & \Omega_{32} & \Omega_{33} & \Omega_{34} \\
    \Omega_{41} & \Omega_{42} & \Omega_{43} & \Omega_{44} \\
  \end{array}
\right]\left[
         \begin{array}{c}
           \mathcal{E}^{or} \\
           \mathcal{E}^{er} \\
           \mathcal{E}^{ot} \\
           \mathcal{E}^{et} \\
         \end{array}
       \right]
       =
       \mathcal{E}^{oi}\left[
                         \begin{array}{c}
                           \Upsilon_{1} \\
                           \Upsilon_{2} \\
                           \Upsilon_{3} \\
                           \Upsilon_{4} \\
                         \end{array}
                       \right]
\nonumber
\end{equation}
\begin{equation}
+
\mathcal{E}^{ei}
\left[
\begin{array}{c}
\Gamma_{1} \\
\Gamma_{2} \\
\Gamma_{3} \\
\Gamma_{4} \\
\end{array}
\right],
\label{omega}
\end{equation}
we can determine the value of the spectral amplitudes from Cramer's rule
\begin{eqnarray}
\mathcal{E}^{or}&=&
\frac{\mathcal{E}^{oi}\det\left[\Omega_{1}^{\Upsilon}\right]+\mathcal{E}^{ei}\det\left[\Omega_{1}^{\Gamma}\right]}
{\det\left[\Omega\right]}\\
\mathcal{E}^{er}&=&
\frac{\mathcal{E}^{oi}\det\left[\Omega_{2}^{\Upsilon}\right]+\mathcal{E}^{ei}\det\left[\Omega_{2}^{\Gamma}\right]}
{\det\left[\Omega\right]}\\
\mathcal{E}^{ot}&=&
\frac{\mathcal{E}^{oi}\det\left[\Omega_{3}^{\Upsilon}\right]+\mathcal{E}^{ei}\det\left[\Omega_{3}^{\Gamma}\right]}
{\det\left[\Omega\right]}\label{Eot}\\
\mathcal{E}^{et}&=&
\frac{\mathcal{E}^{oi}\det\left[\Omega_{4}^{\Upsilon}\right]+\mathcal{E}^{ei}\det\left[\Omega_{4}^{\Gamma}\right]}
{\det\left[\Omega\right]}\label{Eet}
\end{eqnarray}
Where $\Omega$ is the $4\times 4$ matrix of (\ref{omega}) and the $\Omega_{l}^{\Upsilon}$ and $\Omega_{l}^{\Gamma}$ matrices are found by changing the l-th column of the $\Omega$ matrix by the matrix columns $\Gamma$ and $\Upsilon$, respectively.

Knowing the amplitudes of the incident and transmitted fields, we can determine the field at the interface $2-3$
($z=z0+L$),and finally calculate the field in the media $3$. The boundary conditions are analogous, that is,
\begin{equation}
\hat{z}\times\left[\vec{\mathcal{E}}^{i}+\vec{\mathcal{E}}^{r}-\vec{\mathcal{E}}^{t}\right]_{z=z_{0}+L}=0
\end{equation}
and
\begin{equation}
\hat{z}\times\left[\vec{\mathcal{H}}^{i}+\vec{\mathcal{H}}^{r}-\vec{\mathcal{H}}^{t}\right]_{z=z_{0}+L}=0.
\end{equation}
We must calculate the incident angular spectrum at interface $II$, $\mathcal{E}_{II}^{i}$. For this we will define that
\begin{equation}
\hat{s}^{\mu}=\sigma^{\mu}\hat{z}+a^{\mu}\hat{a}+b^{\mu}\hat{b}.
\label{smu}
\end{equation}
If $\mu=i$ or $r$, equation (\ref{smu}) refers to medium $2$, while if $\mu=t$, it refers to medium $3$. Thus,
$\hat{s}^{i}=\hat{s}^{r}=\hat{s}_{2}$, $\hat{s}^{t}=\hat{s}_{3}$ the same holds for $a^{\mu}$, $b^{\mu}$ and $\sigma^{\mu}$.
\subsection{The electric field and the angular spectrum in the medium $2$ and $3$}
The electric field in the medium $2$ is
\begin{equation}
\vec{E}(\vec{r})=\int_{\mathbb{R}^{2}}\mathcal{E}_{I}^{et}\hat{\epsilon}_{e}e^{i\vec{k}^{e}\cdot\vec{r}}d\vec{q}+
\int_{\mathbb{R}^{2}}\mathcal{E}_{I}^{ot}\hat{\epsilon}_{o}e^{i\vec{k}^{o}\cdot\vec{r}}d\vec{q},
\label{espang2}
\end{equation}
where $\hat{\epsilon}_{e}$, $\hat{\epsilon}_{o}$, $k_{z}^{e}$ and $k_{z}^{o}$ are calculated in the medium $2$, and $\vec{k}^{e}$ must be calculated taking into account that the wave propagates in increasing $z$-direction.

According to equations (\ref{Eot}) and (\ref{Eet}), the $\mathcal{E}_{I}^{ot}$ and $\mathcal{E}_{I}^{et}$ amplitudes are
\begin{eqnarray}
\mathcal{E}_{I}^{ot}&=&\mathcal{A}\mathcal{E}_{I}^{oi}+\mathcal{B}\mathcal{E}_{I}^{ei},\\
\mathcal{E}_{I}^{et}&=&\mathcal{C}\mathcal{E}_{I}^{oi}+\mathcal{D}\mathcal{E}_{I}^{ei}
\end{eqnarray}
where
\begin{eqnarray}
\mathcal{A}&=&\det\left[\Omega_{3}^{\Upsilon}\right]/\det\left[\Omega\right]\\
\mathcal{B}&=&\det\left[\Omega_{3}^{\Gamma}\right]/\det\left[\Omega\right]\\
\mathcal{C}&=&\det\left[\Omega_{4}^{\Upsilon}\right]/\det\left[\Omega\right]\\
\mathcal{D}&=&\det\left[\Omega_{4}^{\Gamma}\right]/\det\left[\Omega\right].
\end{eqnarray}
By making $z=L$ in equation (\ref{espang2}), we obtain the angular spectrum of the incident field at interface $II$, $\vec{\mathcal{E}}_{II}^{i}$, that we can apply the same previous procedures to obtain the field in media $3$, $\vec{\mathcal{E}}_{II}^{t}$, as a function of the incident field at interface $I$, $\vec{\mathcal{E}}_{I}^{i}$.
In all the calculations we are neglecting terms of higher order, that is, the field in the media $1$ reflected by the interface $II$ and transmitted by the interface $I$ and the multiple reflections inside the media $2$.

The angular spectrum incident on interface $II$ shall be
\begin{equation}
\vec{\mathcal{E}}_{II}^{i}=\vec{\mathcal{E}}_{II}^{oi}\hat{\epsilon}_{o}+\vec{\mathcal{E}}_{II}^{ei}\hat{\epsilon}_{e}
\end{equation}
where $\hat{\epsilon}_{o}$ and $\hat{\epsilon}_{e}$ correspond to medium $2$ and
\begin{eqnarray}
\mathcal{E}_{II}^{oi}&=&\mathcal{E}_{I}^{ot}e^{ik_{z}^{o}L}=\left(\mathcal{A}\mathcal{E}_{I}^{oi}+
\mathcal{B}\mathcal{E}_{I}^{ei}\right)e^{ik_{z}^{o}L}\label{espIIoi}\\
\mathcal{E}_{II}^{ei}&=&\mathcal{E}_{I}^{et}e^{ik_{z}^{e}L}=\left(\mathcal{C}\mathcal{E}_{I}^{oi}+
\mathcal{D}\mathcal{E}_{I}^{ei}\right)e^{ik_{z}^{e}L}\label{espIIei}
\end{eqnarray}
Through the transfer matrix and using the same procedure used to calculate the fields transmitted to the media $2$, we can calculate the field transmitted to the media $3$. The angular spectra transmitted through interface $II$ will be
\begin{eqnarray}
\mathcal{E}_{II}^{ot}&=&\mathcal{F}\mathcal{E}_{II}^{oi}+\mathcal{G}\mathcal{E}_{II}^{ei},\label{espIIot}\\
\mathcal{E}_{II}^{et}&=&\mathcal{L}\mathcal{E}_{II}^{oi}+\mathcal{M}\mathcal{E}_{II}^{ei}\label{espIIet}
\end{eqnarray}
where
\begin{eqnarray}
\mathcal{F}&=&\det\left[\Lambda_{3}^{\Upsilon}\right]/\det\left[\Lambda\right]\\
\mathcal{G}&=&\det\left[\Lambda_{3}^{\Gamma}\right]/\det\left[\Lambda\right]\\
\mathcal{L}&=&\det\left[\Lambda_{4}^{\Upsilon}\right]/\det\left[\Lambda\right]\\
\mathcal{M}&=&\det\left[\Lambda_{4}^{\Gamma}\right]/\det\left[\Lambda\right].
\end{eqnarray}
The $\Lambda$ matrix has expression identical to expression (\ref{matrixtrans}) for $\Omega$, however, taking into account that in the expressions for $A^{pm}$, $B^{pm}$ and $D^{pm}$,
the indexes $m=i$ and $m=r$ indicate that the quantities should be calculated in the media $2$ (uniaxial), while $m=t$ refers to the media $3$ (isotropic).

Since media $1$ and $3$ are isotropic, the choice of $\hat{s}_{1}$ and $\hat{s}_{3}$ is arbitrary. For example, we can make $\hat{s}_{1}=\hat{s}_{3}=\hat{z}$. This choice simplifies the matrices $\Omega$ and $\Lambda$, since it will do $A^{or}=A^{er}=A^{oi}=A^{ei}=1$, $D^{or}=D^{er}=D^{oi}=D^{ei}=0$ in $\Omega$, and $A^{ot}=A^{et}=1$,$D^{ot}=D^{et}=0$ in $\Lambda$. However, depending on the polarization state of the incident beam, other choices for $\hat{s}_{1}$ and $\hat{s}_{3}$ may be more convenient. Substituting (\ref{espIIoi}) and (\ref{espIIei}) in (\ref{espIIot}) and (\ref{espIIet}), we arrive at a transfer matrix from media $1$ to media $3$ at the base and $\hat{\epsilon}_{o}$ and $\hat{\epsilon}_{e}$:
\begin{equation}
\left(
  \begin{array}{c}
    \mathcal{E}_{II}^{ot} \\
    \mathcal{E}_{II}^{et} \\
  \end{array}
\right)=
\left(
  \begin{array}{cc}
    T^{oo} & T^{oe} \\
    T^{eo} & T^{ee} \\
  \end{array}
\right)
\left(
  \begin{array}{c}
    \mathcal{E}_{I}^{oi} \\
    \mathcal{E}_{I}^{ei} \\
  \end{array}
\right),
\label{matrixT}
\end{equation}
where
\begin{eqnarray}
T^{oo}&=&\mathcal{A}\mathcal{F}e^{ik_{z}^{o}L}+\mathcal{C}\mathcal{G}e^{ik_{z}^{e}L},\\
T^{oe}&=&\mathcal{B}\mathcal{F}e^{ik_{z}^{o}L}+\mathcal{D}\mathcal{G}e^{ik_{z}^{e}L},\\
T^{eo}&=&\mathcal{A}\mathcal{L}e^{ik_{z}^{o}L}+\mathcal{C}\mathcal{M}e^{ik_{z}^{e}L},\\
T^{ee}&=&\mathcal{B}\mathcal{L}e^{ik_{z}^{o}L}+\mathcal{D}\mathcal{M}e^{ik_{z}^{e}L}.
\end{eqnarray}
We will call the matrix of equation (\ref{matrixT}) of matrix T, that is,
\begin{equation}
T=\left(
  \begin{array}{cc}
    T^{oo} & T^{oe} \\
    T^{eo} & T^{ee} \\
  \end{array}
\right).
\end{equation}

\section{Results and discussion}
\begin{figure}
\begin{center}
  \includegraphics[width=5.0cm]{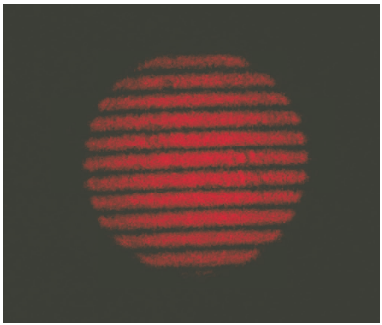}\\
  \caption{Profile of a Gaussian beam with $\lambda=632.8$ nm after crossing a uniaxial crystal.}\label{fotoperfil}
\end{center}
\end{figure}
\begin{figure}
\begin{center}
  \includegraphics[width=7.0cm]{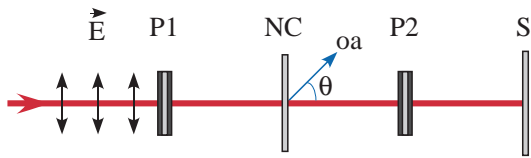}\\
  \caption{An $\vec{E}$ field, with vertical polarization, crosses polarizer P$1$, polarized at $45^{\circ}$, and then passes through the nonlinear crystal NC. The optical axis of the crystal, oa, is in a plane formed by the direction of propagation and the direction perpendicular to propagation. The transmitted beam passes through polarizer P$2$ and the beam image is seen in S.}
\label{interfcnl}
\end{center}
\end{figure}

To show that the equations here described correctly describe the propagation of the fields through a uniaxial crystal of length L,we will make the comparison between the theoretical and experimental transversal profile of a beam when passing through these media. Figure \ref{fotoperfil} shows the intensity profile photograph of a Gaussian beam of wavelength $632.8$ nm when it passes through a Type $I$ uniaxial $BBO$ crystal of length $2.0$ mm.
This image is obtained by assembling the experimental apparatus shown in figure \ref{interfcnl}.
\begin{figure}
\begin{center}
  \includegraphics[width=7.0cm]{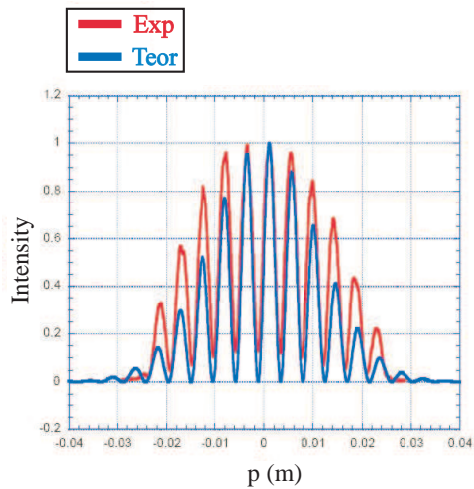}\\
  \caption{
Intensity profile a Gaussian beam with $\lambda=632.8$ nm after passing through a uniaxial crystal. The beam is shown in the figure \ref{fotoperfil}. The horizontal axis is the p position. The interference fringes are approximately in agreement between the experimental (red line) and theoretical (blue line) curves.}
\label{curvas}
\end{center}
\end{figure}
In the figure, an electric field of vertical polarization $\vec{E}$ passes through the polarizer $P1$ polarized at $45^{\circ}$ with respect to the incident beam. Light polarized at $45^{\circ}$ passes through a negative nonlinear crystal ($n_{e} < n_{o}$) whose optical axis lies in the plane formed by the propagation direction of the beam and the vertical direction. When crossing the crystal there is a delay of the extraordinary component with respect to the ordinary component. To observe the interference between the extraordinary and ordinary fields, we need to combine them through the polarizer $P2$ that is polarized identically to $P1$. In Figure \ref{curvas} we have the curve coming from equation (\ref{matrixT}) compared to the maximum and minimum of figure \ref{fotoperfil}.
\begin{figure}
\begin{center}
  \includegraphics[width=6.0cm]{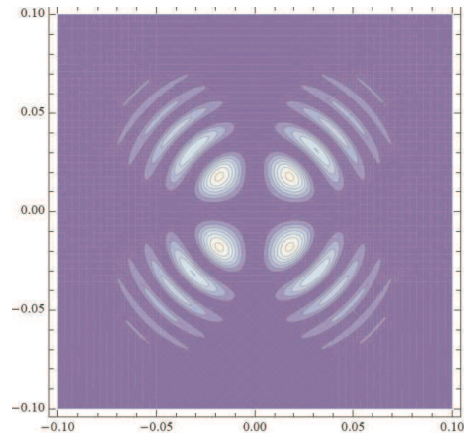}\\
  \caption{Conoscopic figure of a Gaussian profile laser with $\lambda=632.8$ nm and total divergence $2\phi=0.2$ rad, traversing a uniaxial crystal $ BBO $ of length $L=10$ mm in the direction of the optical axis.}
\label{conoscopic}
\end{center}
\end{figure}
We see that the periods of the theoretical and experimental curves agree satisfactorily. Another proof of expression (\ref{matrixT}) is seen through the conoscopic figure \cite{yu} shown in Figure \ref{conoscopic}. The conoscopic figure is obtained through equation (\ref{matrixT}) when the incident beam has linear polarization at $45^{\circ}$ with the x-axis of figure \ref{cris} and is highly focused. The crystal is a $BBO$ of length $L=10$ mm crossed by a laser of wavelength $\lambda= 632.8$ nm and divergence $2\varphi=0.2$ rad.
The light passes through the crystal toward the optical axis. Equation (\ref{matrixT}) shows the vector character of the angular spectrum in anisotropic media, since the $T^{oe}$ and $T^{eo}$ elements couple the two orthogonal components of polarization. The scalar approximation will be valid when the $T^{oe}$ and $T^{eo}$ elements can be neglected.
\begin{figure}
\center
\subfigure[ref1][ Dependence of the $T^{oo}$ with $q_{x}$ and $q_{y}$ when $\theta=45^{\circ}$]
{\includegraphics[width=5cm]{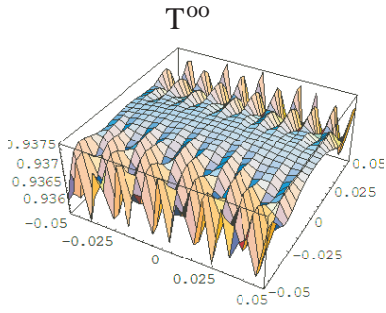}}
\qquad
\subfigure[ref2][ Dependence of the $T^{ee}$ with $q_{x}$ and $q_{y}$ when $\theta=45^{\circ}$]
{\includegraphics[width=5cm]{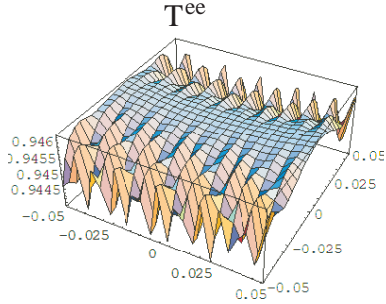}}
\qquad
\subfigure[ref3][ Dependence of the $T^{oe}$ with $q_{x}$ and $q_{y}$ when $\theta=45^{\circ}$]{\includegraphics[width=5cm]{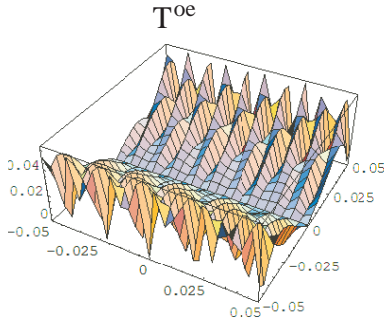}}
\qquad
\subfigure[ref4][ Dependence of the $T^{eo}$ with $q_{x}$ and $q_{y}$ when $\theta=45^{\circ}$]{\includegraphics[width=5cm]{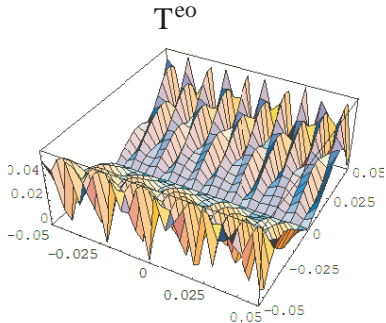}}
\caption{\label{fig:tqc}Dependence of the quantities $T^{oo}$, $T^{ee}$, $T^{oe}$ and $T^{eo}$ with $q_{x}$ and $q_{y}$ when $\theta=45^{\circ}$. All functions are normalized with the maximum value being equal to $1$. We see that for $T^{oo}$ there is a small variation around $0.937$, that is, the function varies very little for this angle and can be considered as a constant. The same is true for $T^{ee}$ which has a small variation around $0.945$. We also observed that the  $T^ {oe}$ and $T^{eo}$ quantities are practically zero for $\theta=45^{\circ}$.
}
\end{figure}
\begin{figure}
\center
\subfigure[ref1][ Dependence of the $T^{oo}$ with $q_{x}$ and $q_{y}$ when $\theta=90^{\circ}$]{\includegraphics[width=5cm]{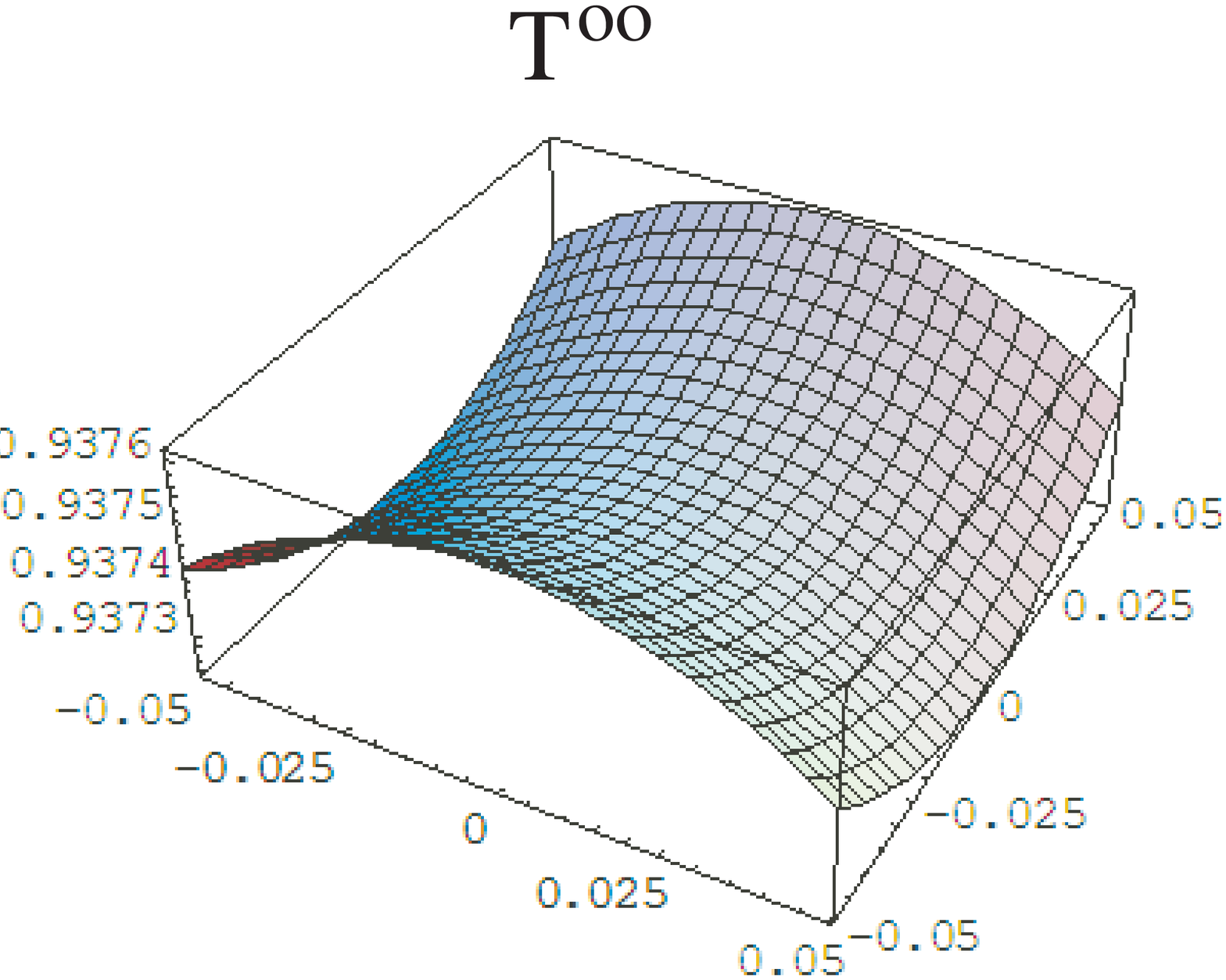}}
\qquad
\subfigure[ref2][ Dependence of the $T^{ee}$ with $q_{x}$ and $q_{y}$ when $\theta=90^{\circ}$]{\includegraphics[width=5cm]{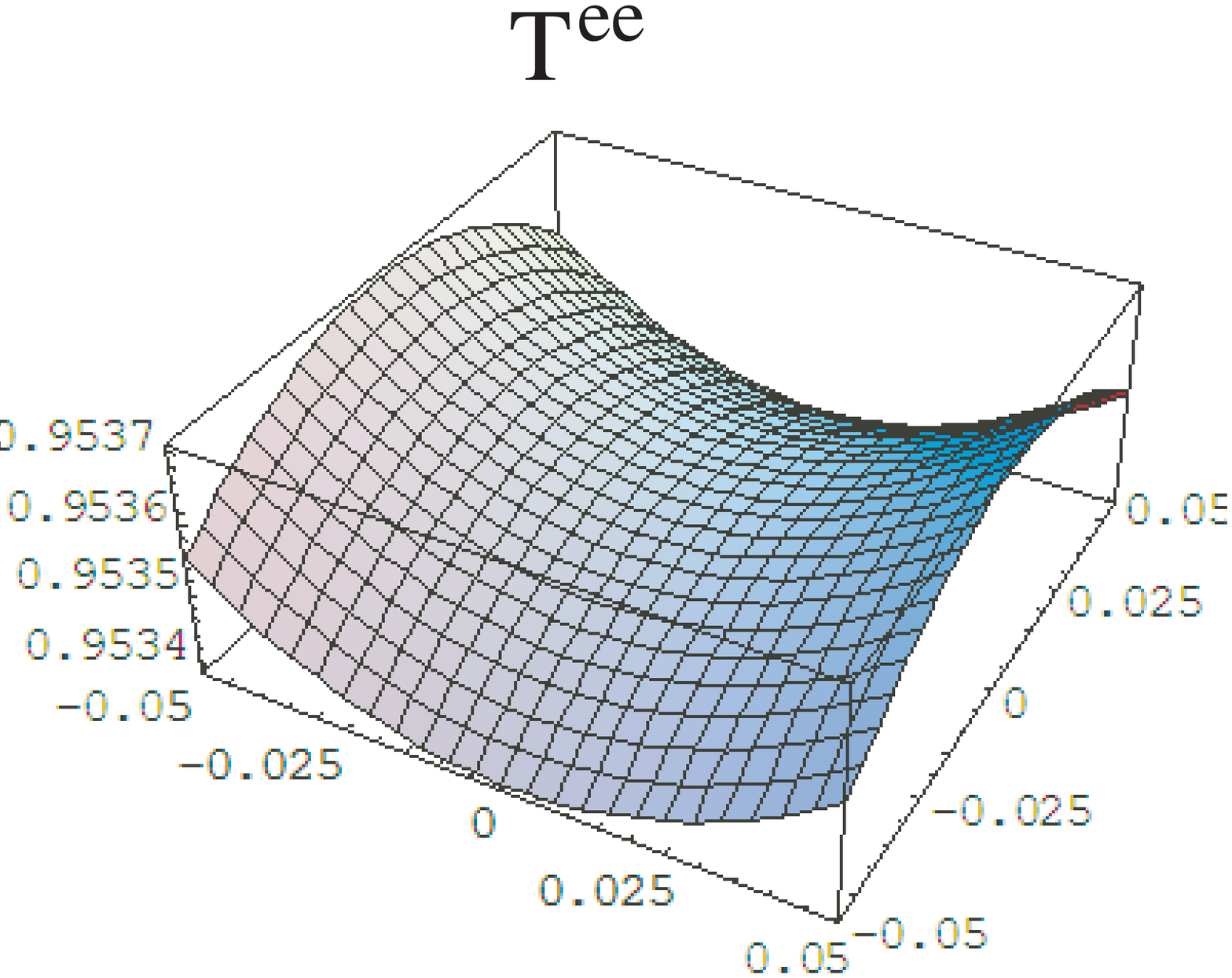}}
\qquad
\subfigure[ref3][ Dependence of the $T^{oe}$ with $q_{x}$ and $q_{y}$ when $\theta=90^{\circ}$]{\includegraphics[width=5cm]{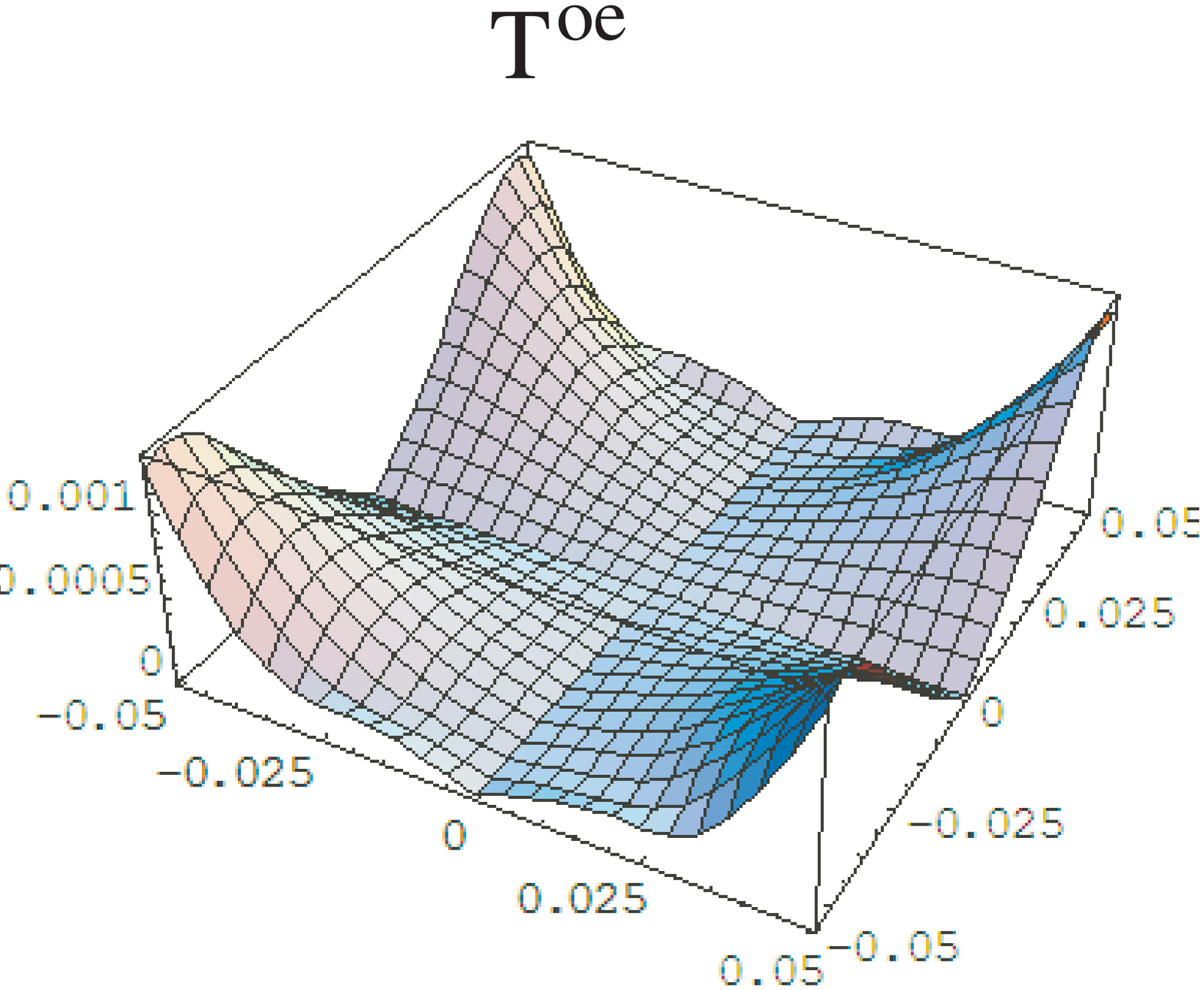}}
\qquad
\subfigure[ref4][ Dependence of the $T^{eo}$ with $q_{x}$ and $q_{y}$ when $\theta=90^{\circ}$]{\includegraphics[width=5cm]{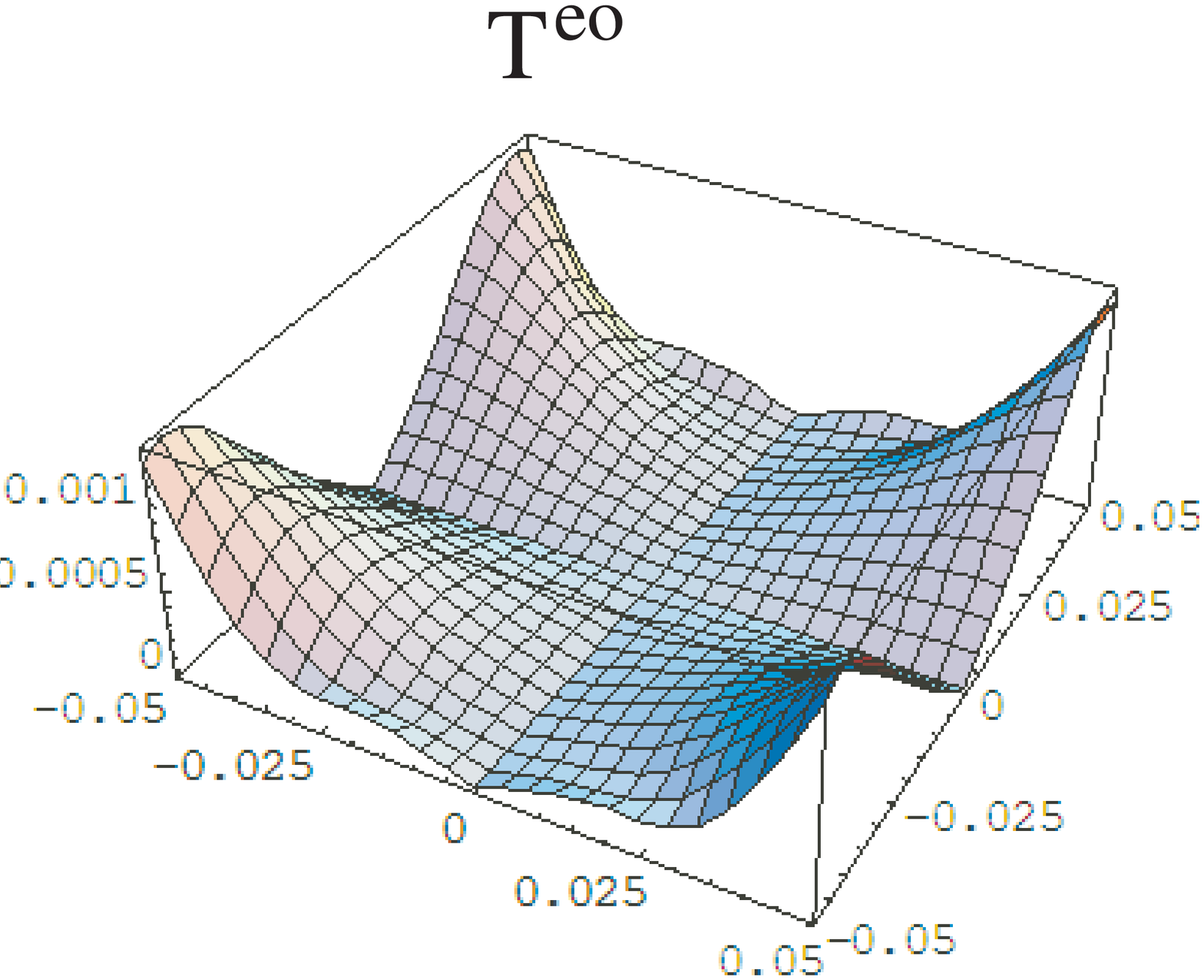}}
\caption{\label{fig:tnv}Dependence of the quantities $T^{oo}$, $T^{ee}$, $T^{oe}$ and $T^{eo}$ with $q_{x}$ and $q_{y}$ when $\theta=90^{\circ}$. All functions are normalized with the maximum value being equal to $1$. We see that for $T^{oo}$ there is a small variation around $0.937$, that is, the function varies very little for this angle and can be considered as a constant. The same is true for $T^{ee}$ which has a small variation around $0.953$. We also observed that the  $T^ {oe}$ and $T^{eo}$ quantities are practically zero for $\theta=90^{\circ}$.}
\end{figure}
\begin{figure}
\center
\subfigure[ref1][ Dependence of the $T^{oo}$ with $q_{x}$ and $q_{y}$ when $\theta=0^{\circ}$]{\includegraphics[width=5cm]{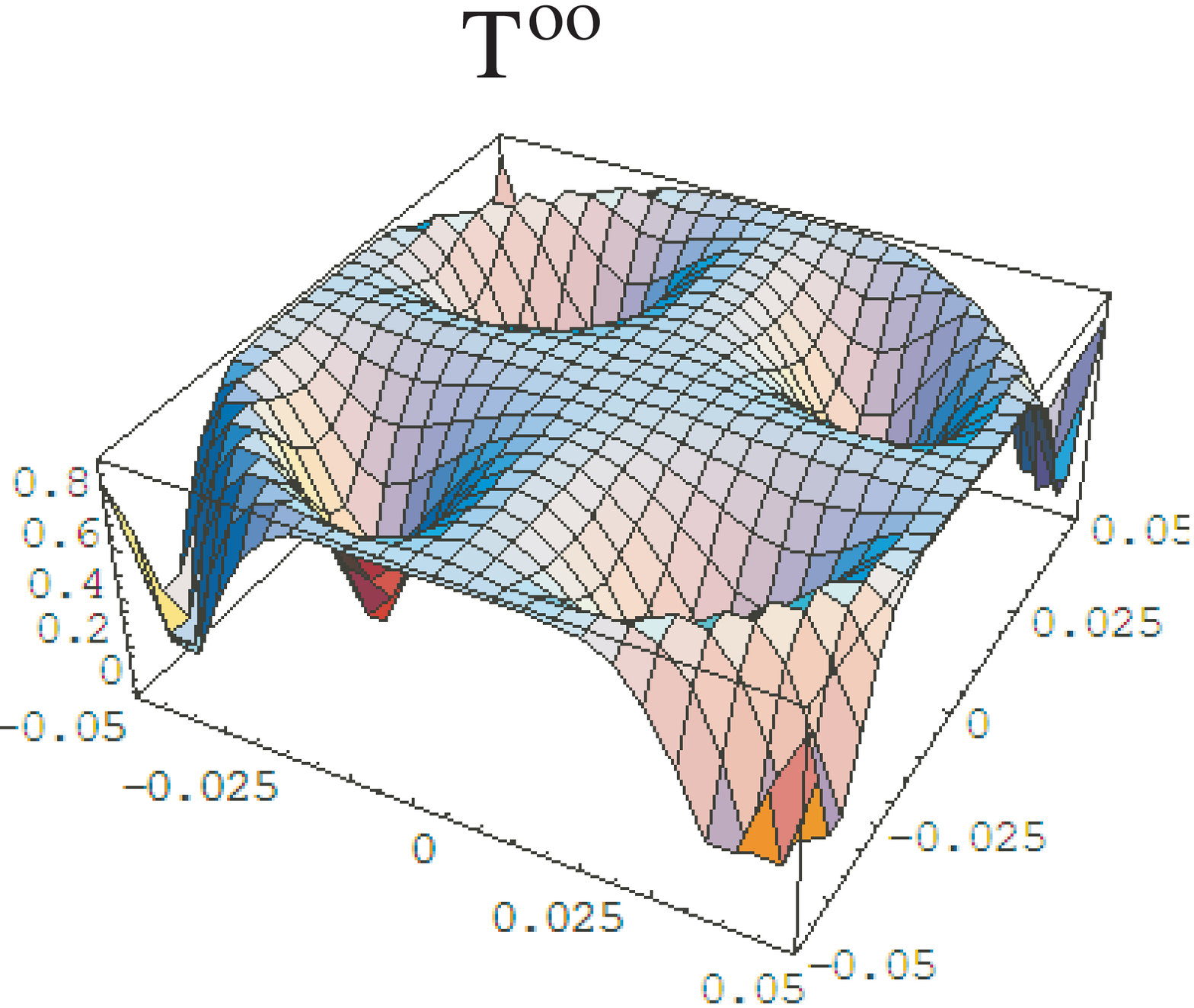}}
\qquad
\subfigure[ref2][ Dependence of the $T^{ee}$ with $q_{x}$ and $q_{y}$ when $\theta=0^{\circ}$]{\includegraphics[width=5cm]{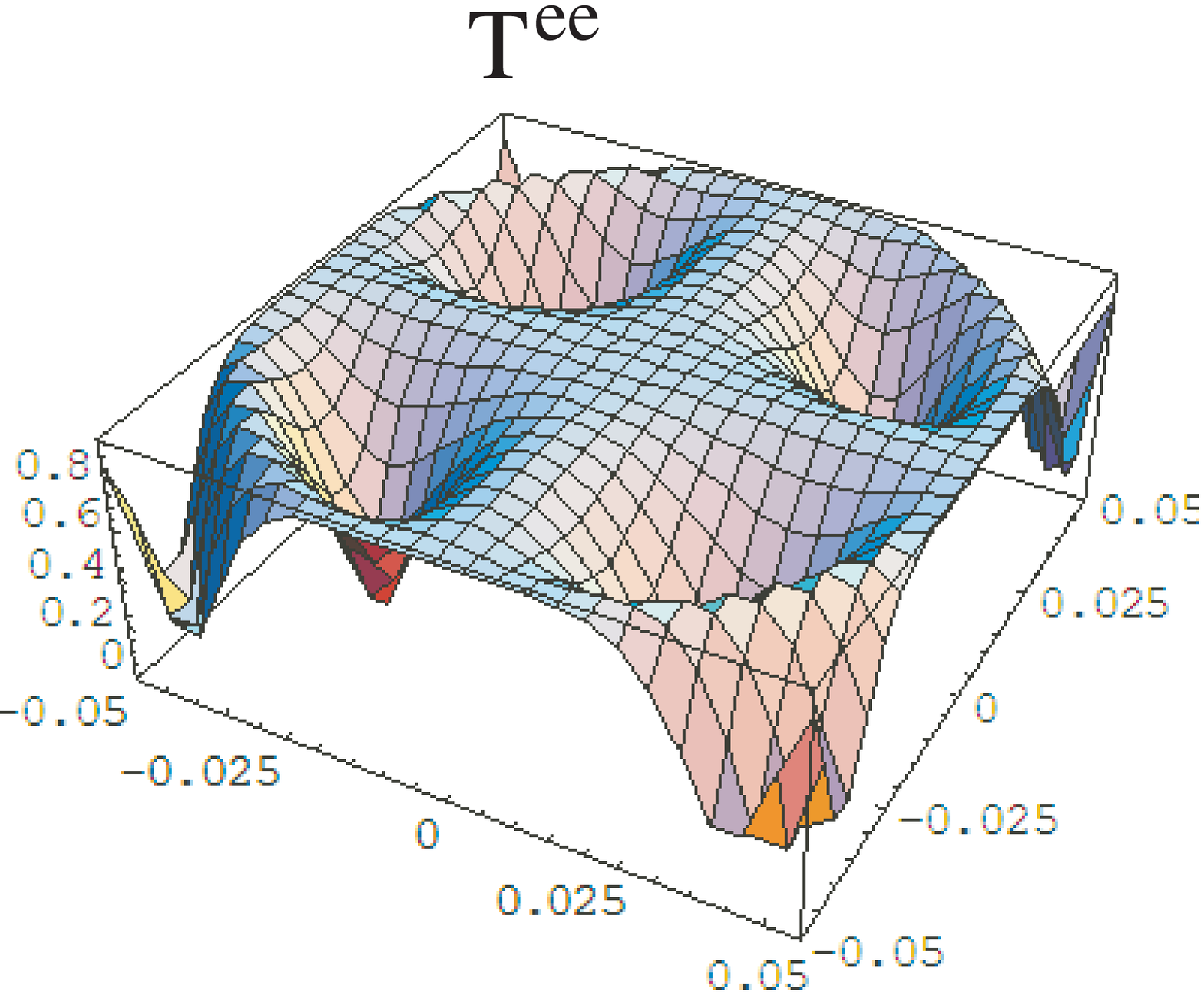}}
\qquad
\subfigure[ref3][ Dependence of the $T^{oe}$ with $q_{x}$ and $q_{y}$ when $\theta=0^{\circ}$]{\includegraphics[width=5cm]{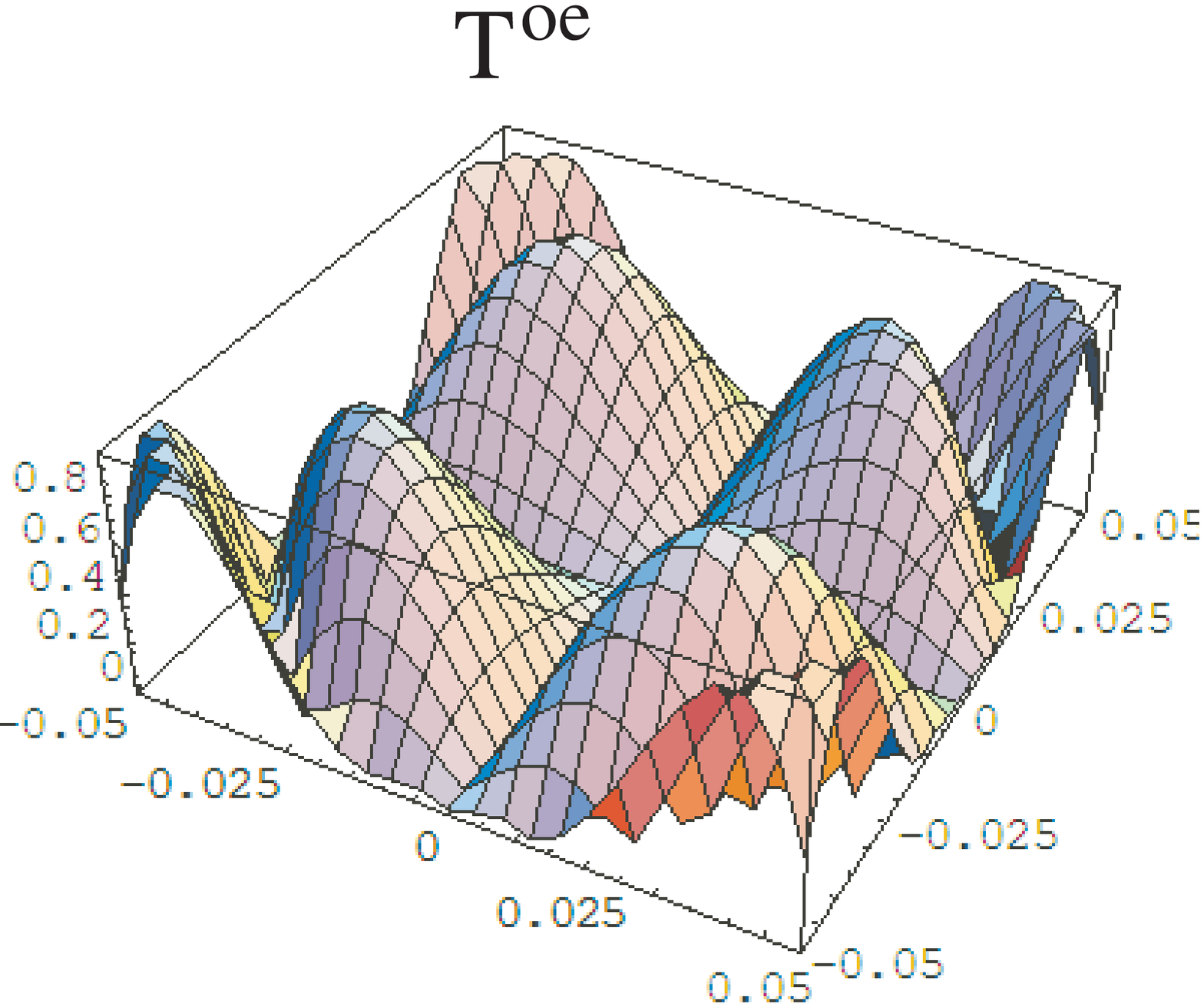}}
\qquad
\subfigure[ref4][ Dependence of the $T^{eo}$ with $q_{x}$ and $q_{y}$ when $\theta=0^{\circ}$]{\includegraphics[width=5cm]{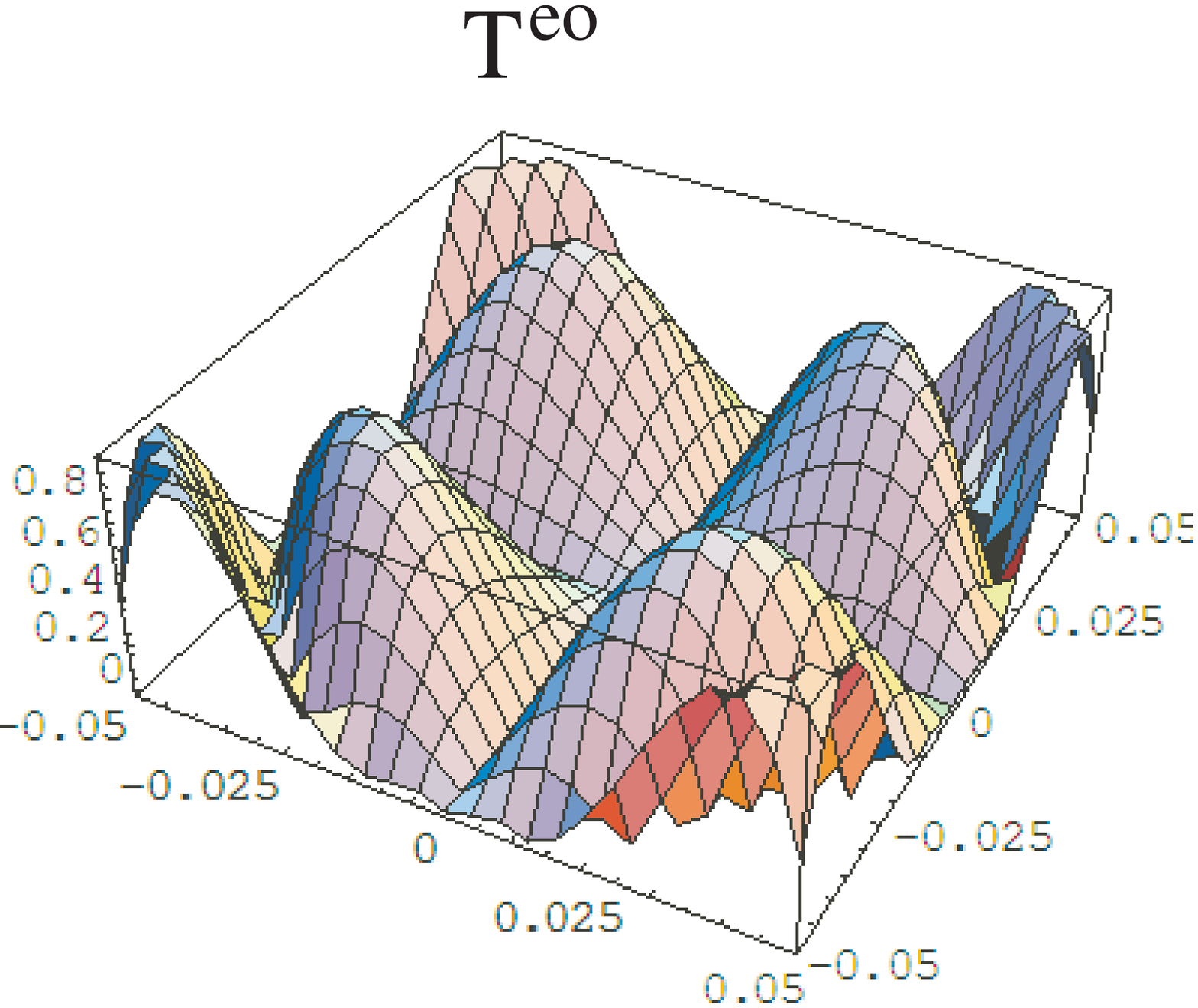}}
\caption{\label{fig:tze}Dependence of the quantities $T^{oo}$, $T^{ee}$, $T^{oe}$ and $T^{eo}$ with $q_{x}$ and $q_{y}$ when $\theta=0^{\circ}$. All functions are normalized with the maximum value being equal to $1$. We see that the functions $T^{oo}$, $T^{ee}$,  $T^{oe}$ and $T^{eo}$ vary strongly with $q_{x}$ e $q_{y}$ for $\theta=0^{\circ}$.}
\end{figure}
Figures \ref{fig:tqc}, \ref{fig:tnv} and \ref{fig:tze} show how the quantities $T^{oo}$, $T^{ee}$, $T^{eo}$ and $T^{oe}$ vary as a function of $q_{x}$ and $q_{y}$ for certain angles $\theta$ of the optical axis. In Figure \ref{fig:tqc} we do $\theta=45^{\circ}$ and we observe that these functions vary very little around a certain value so that we can consider them as constants. This behavior is also observed for angles that are close to $45^{\circ}$, and are usually angles that allow the phase matching for the parametric down conversion \cite{mandel}. We note that the terms $T^{eo}$ and $T^{oe}$ are practically null and that the terms $T^{oo}$ and $T^{ee}$ are very close to $1$.
These terms are like transmission coefficients of the fields and they do not reach the unit for the account of the reflection in the interfaces.
In Figure \ref{fig:tnv} we show the variations of the quantities $T^{oo}$, $T^{ee}$, $T^{eo}$ and $T^{oe}$ as a function of $q_{x}$ and $q_{y}$, for $\theta=90^{\circ}$. We again see that we can consider these functions as constants, with $T^{oe}$ and $T^{eo}$ negligible. In these two cases, where we can neglect $T^{oe}$ and $T^{eo}$, each component of the angular spectrum can be treated independently.
The situation changes completely when we do $\theta=0^{\circ}$. Figure \ref{fig:tze} shows that $T^{oo}$, $T^{ee}$,  $T^{oe}$ and $T^{eo}$ vary greatly with $q_{x}$ and $q_{y}$, which makes it impossible to give a scalar treatment for propagation.
In this case there is coupling between the ordinary and extraordinary fields within the birefringent medium, that is, if the incident field has extraordinary polarization, for example, the output field will have the extraordinary and ordinary components. The ordinary component can not be neglected and the vector formulation is necessary to explain the propagation of the fields.
Equation (\ref{matrixT}) allows to determine the angular spectrum in media $3$ by knowing the incident angular spectrum, coming from media $1$.
The determination of the angular spectrum in the media $3$ of the beams traversing uniaxial crystals allows to analyze the process of propagation of any electromagnetic beam in these crystals. The knowledge of the transfer matrix allows a study of several interesting phenomena, among them, a detailed analysis of the orbital angular momentum of the Laguerre-Gaussian beams \cite{allen, allencour, beijer}. These have phase singularities that can be conserved during the \cite{walborn} propagation process. With the determination of the field in the media $3$ (crystal exit) we can analyze in detail the propagation of these beams not only along the optical axis or in the direction perpendicular to it, but also make an analysis along any direction.
\section{Conclusions}
We present the equations of the k-surfaces, equations (\ref{eq:esfera}) and (\ref{eq:elipse}), where we obtain the effects of anisotropy on uniaxial media given by (\ref{kno})-(\ref{gamma}). The field within the crystal can be written on the basis of ordinary and extraordinary polarization vectors and the plane waves of the angular spectrum will have two distinct polarization components. We determine the transfer matrix, represented by equation (\ref{matrixtrans}), which relates the fields between two medias. This enabled a way to determine the angular spectrum at the output of the crystal by knowing the incident angular spectrum. We observe the behavior of the quantities $T^{oo}$, $T^{ee}$, $T^{eo}$ and $T^{oe}$ as a function of $q_{x}$ and $q_{y}$ and analyze them in which cases we can give a scalar or vector treatment for the propagation of the electromagnetic beams in uniaxial media.
\section{Acknowledgments}
This work was supported by CNPq -- Conselho Nacional de Desenvolvimento
Cient\'{\i}fico e Tecnol\'ogico, Instituto do Mil\^enio de
Informa\c{c}\~ao Qu\^antica, CAPES and FAPEMIG.

\end{document}